\documentclass{PoS}

\newcommand{\be}{\begin{equation}}
\newcommand{\ee}{\end{equation}}
\newcommand{\ba}{\begin{eqnarray}}
\newcommand{\ea}{\end{eqnarray}}
\newcommand{\bi}{\begin{itemize}}
\newcommand{\ei}{\end{itemize}}

\title{
\vspace*{-3.5cm}
\begin{flushright}\texttt{\footnotesize
BI-TP 2009/02\\}
\end{flushright}
\vfill
Charmonium correlators and spectral functions at finite temperature }

\ShortTitle{charmonium correlators and spectral functions at finite temperature }

\author{\speaker{H.-T. Ding}, O. Kaczmarek, F. Karsch and H. Satz %
\\
Fakult\"{a}t f\"{u}r Physik,
 Universit\"{a}t Bielefeld, D-33615 Bielefeld, Germany\\
     E-mail: \email{hengtong.ding,okacz,karsch,satz@physik.uni-bielefeld.de}}

\abstract{
We present an operational approach to address the in-medium behavior of charmonium and analyze the reliability of maximum entropy method (MEM). We study the dependences of the ratio of correlators to the reconstructed one and the free one on the resonance's width and the continuum's threshold. Furthermore, we discuss the issue of the default model dependence of the spectral function obtained from MEM. 

}

\FullConference{8th Conference Quark Confinement and the Hadron Spectrum \\
                 September 1-6, 2008\\
                 Mainz, Germany}

\usepackage{epsfig}
\usepackage{graphicx}
\usepackage{dcolumn}
\usepackage{bm}

\begin{document}
\section{Introduction}

Matsui and Satz~\cite{Satz86} proposed long time ago the melting of charmonium states due to color screening can signal the formation of Quark Gluon Plasma in heavy ion collisions.  The suppression of the electrons from $J/\psi$ is observed both at RHIC and SPS~\cite{PHENIX08}, however,  its interpretation is not quite understood.  This phenomenon has been studied quite extensively in potential models~\cite{Mocsy08}, based either on models or finite temperature lattice QCD results for the heavy quark potential in a non-relativistic Schr\"odinger equation. Depending on the form of the heavy quark potential used, the dissociation temperature of charmonium can be ranged from shortly above T$_{c}$ up to values similar to those obtained in lattice QCD~\cite{latticeQCDstudy}.  With the lattice QCD approach, the properties of the charmonium, which can be directly seen from the spectral function, is enclosed in the lattice QCD calculated euclidean time correlation functions. To extract the spectral functions from the correlation function, Maximum Entropy Method~\cite{Asakawa01} is normally used.
  
Here we contribute an operational approach to address the in-medium behavior of charmonium and address the issue of the default model dependence of the spectral function obtained from MEM. 
\section{Lattice correlators and spectral functions}

We look into the momentum-projected Matsubara correlators
\be
G(\tau,\vec{p},T)=\sum_{\vec{x}} e^{i\vec{p}\cdot \vec{x}} <J_{H}(\tau,\vec{x})J_{H}^{\dag}(0,\vec{0})>_T ,
\ee
where $J_H$ is a suitable mesonic operator, $\vec{p}$ is the spatial momentum, T is the temperature of the gluonic plasma and the Euclidean time $\tau\in[0,1/T)$. Through analytic calculation, the Matsubara correlator can be related to the hadronic spectral function as the following:
\be
G_{H}(\tau,\vec{p})=\int_0^{\infty}{\mathrm{d}\omega\sigma_{H}(\omega,\vec{p},T)}\frac{cosh(\omega(\tau-\frac{1}{2T}))}{sinh(\frac{\omega}{2T})},
\ee

Extracting the spectral function at finite temperature lattice QCD is hampered mainly by two issues: the physical extent of time is restricted by the temperature, $\tau<1/T$, and the finite number of correlator points\footnote{what's more, the correlator points are not precise but with statistical errors} making the inversion of Eq.~2.2 ill-posed.

\section{Charmonium correlators}

First, we analyze the sensitivity of the correlators to the spectral function by using the two following references correlators:
\ba
G_{0}(\tau,T)= \int_{0}^{\infty}{\mathrm{d}\omega\sigma(\omega,T=0)K(\tau,T)},~~~~
G_{free}(\tau,T)=\int_{0}^{\infty}{\mathrm{d}\omega\sigma_{free}(\omega,T)K(\tau,T)},
 \ea
where $G_{0}(\tau,T)$ is so called "reconstructed" correlator, and $G_{free}(\tau,T)$ is the free correlator at finite T. 
 $G_{0}(\tau,T)$  and $G_{free}(\tau,T)$ show the behavior if the spectral function at temperature T were identical to that at T=0 and at free limit, respectively. The ratios of the finite T correlators to these two references are:
\be
R_{0}=\frac{G(\tau,T)}{G_{0}(\tau,T)}, ~~~~~~
R_{free}=\frac{G(\tau,T)}{G_{free}(\tau,T)}.
\ee
then $R_{0}\approx 1$ and $R_{free}\approx 1$ indicate the bound state still exist and already melt, respectively.

We consider the spectral function as a combination of the resonance and the continuum: $\sigma=\sigma_{r} + \sigma_{c}$. For the spectral function of the resonance, the following form is taken for T=0:
$\sigma_{r}(\omega,T=0)=\delta(\omega-M)$,
where M donates the mass of resonance. And at finite temperature T, we take the spectral function of the resonance to have the relativistic Breit-Wigner form of
$\sigma_{r}(\omega,\gamma)=N(\gamma)\frac{M}{\pi}\{\frac{2\omega\gamma}{\omega^2\gamma^2+(\omega^2-M^2)^2}\}$,
where $\gamma$ is the width of the resonance, and N($\gamma$) is the normalization factor to maintain the relativistic Breit-Wigner the same strength as the delta function.
For the continuum part of the spectral function, we take the formula of 
$\sigma_{c}=\frac{3}{8\pi^2}\omega^2 tanh(\frac{\omega}{4T})\sqrt{1-(\frac{s}{\omega})^2}~(2+(\frac{s}{\omega})^2)$,
where s is threshold of the continuum, for T=0, $s=s_{0}=4.5$ GeV, for the free case, s=2m (m is mass of the quark),
for finite T, s is T dependent.

The ratio $R_{0}$ is shown in the Fig. 1, in which the left plot is with different values of the resonance's width and the right plot is with different values of the continuum's threshold. We can see that both increasing the resonance's width and decreasing the continuum's threshold can make $R_0$ go farther away from the unity. The influence of the resonance, with the width of 0.9 GeV only making a deviation of 9\% at the symmetry point, is much smaller than that of the continuum, with threshold being 0.8 GeV smaller than $s_{0}$=4.5 GeV making a difference of 20\%. The ratio $R_{free}$ with different  values of the width of resonance and different values of the threshold of the continuum is shown in the left and right plot of Fig. 2, respectively. Similar to $R_{0}$, the influence of the resonance is smaller than that of the continuum.
\begin{figure}[t]
\centering
  \epsfig{file=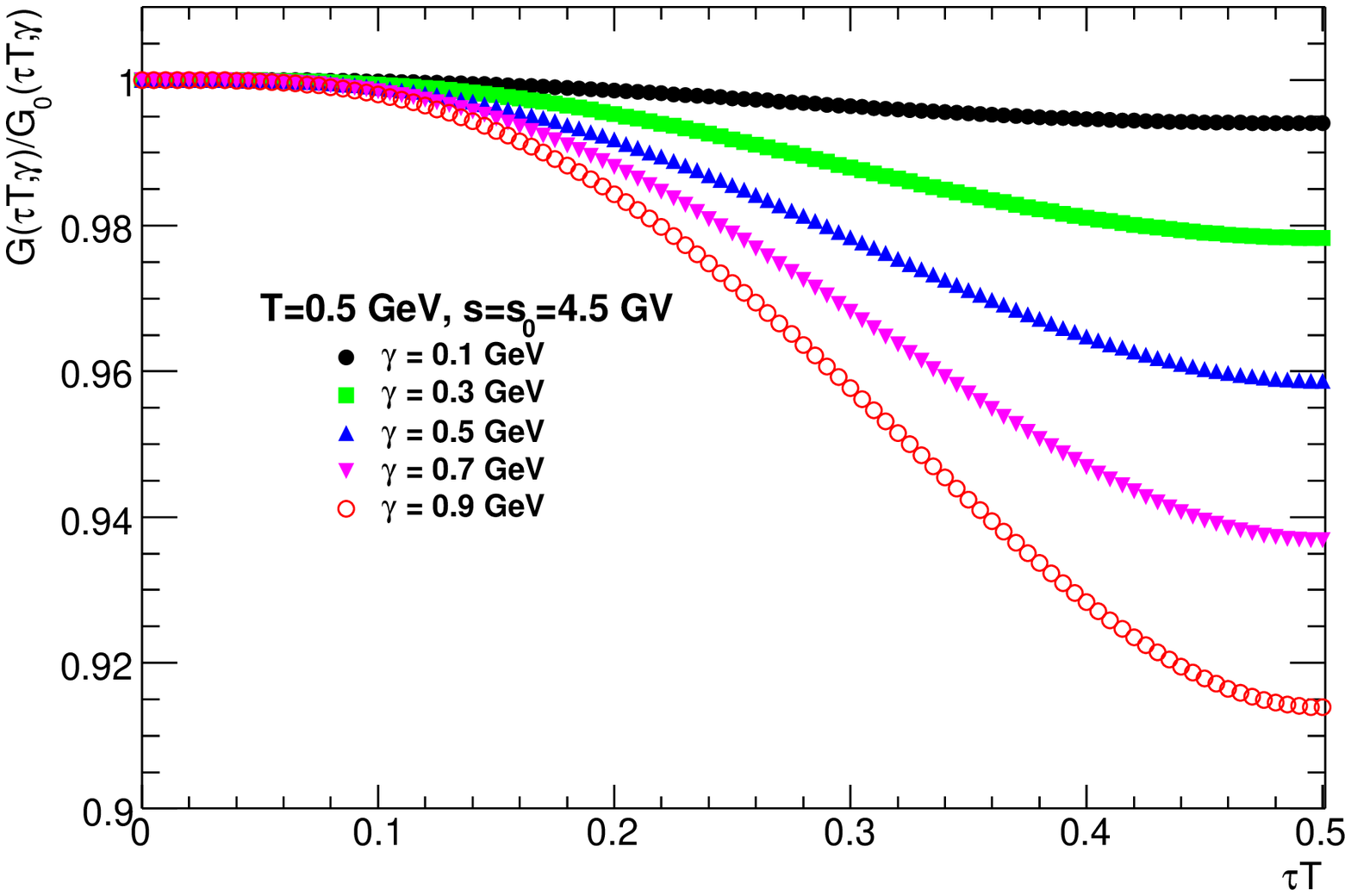,width=0.4\textwidth}
  \epsfig{file=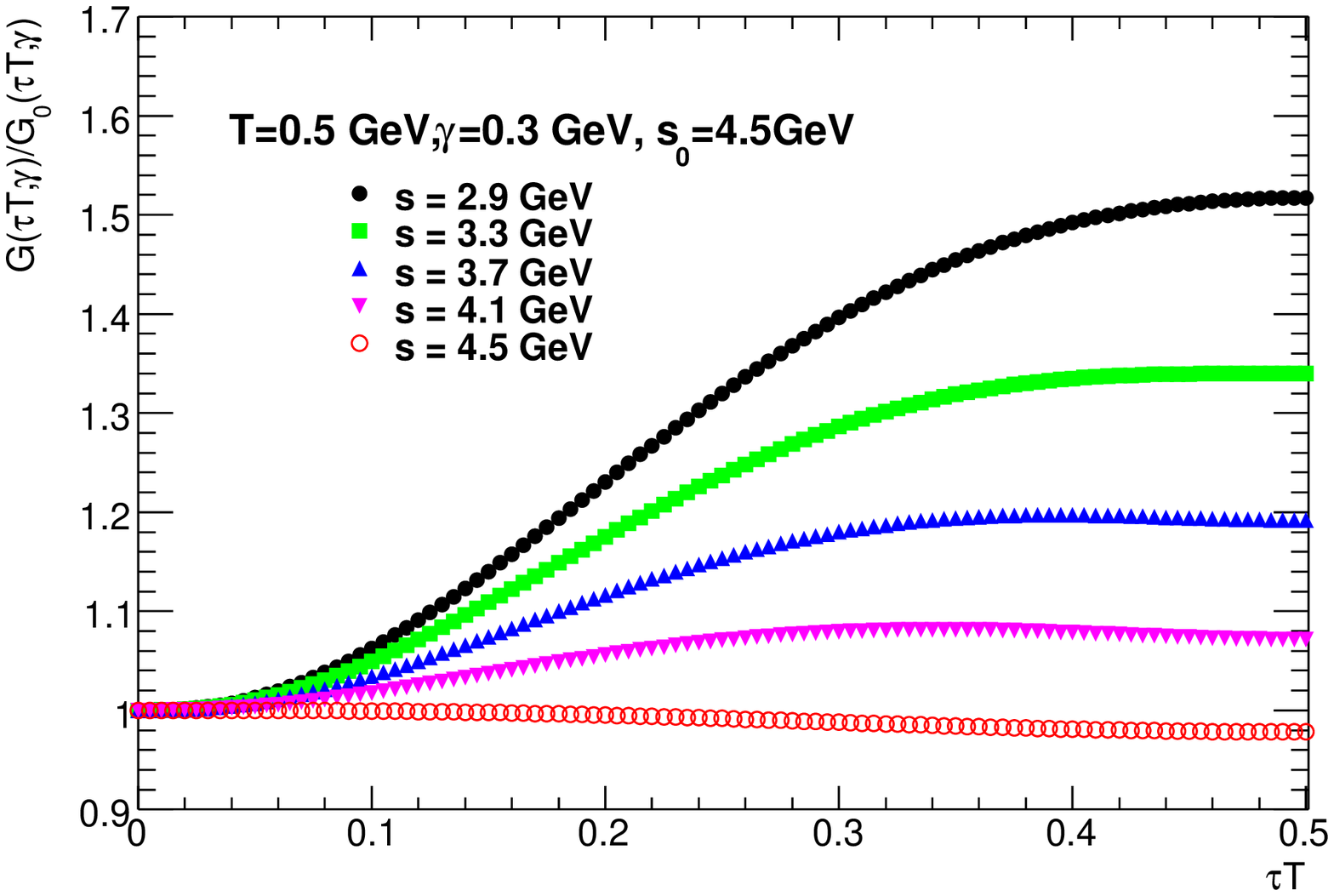,width=0.4\textwidth}
\caption{
Ratio of correlator $G(\tau,T)$ to $G_{0}(\tau,T)$ versus $\tau T$:  with different width of the resonance (left) and different threshold of the continuum (right).
}
\end{figure}
\begin{figure}[t]
\centering
  \epsfig{file=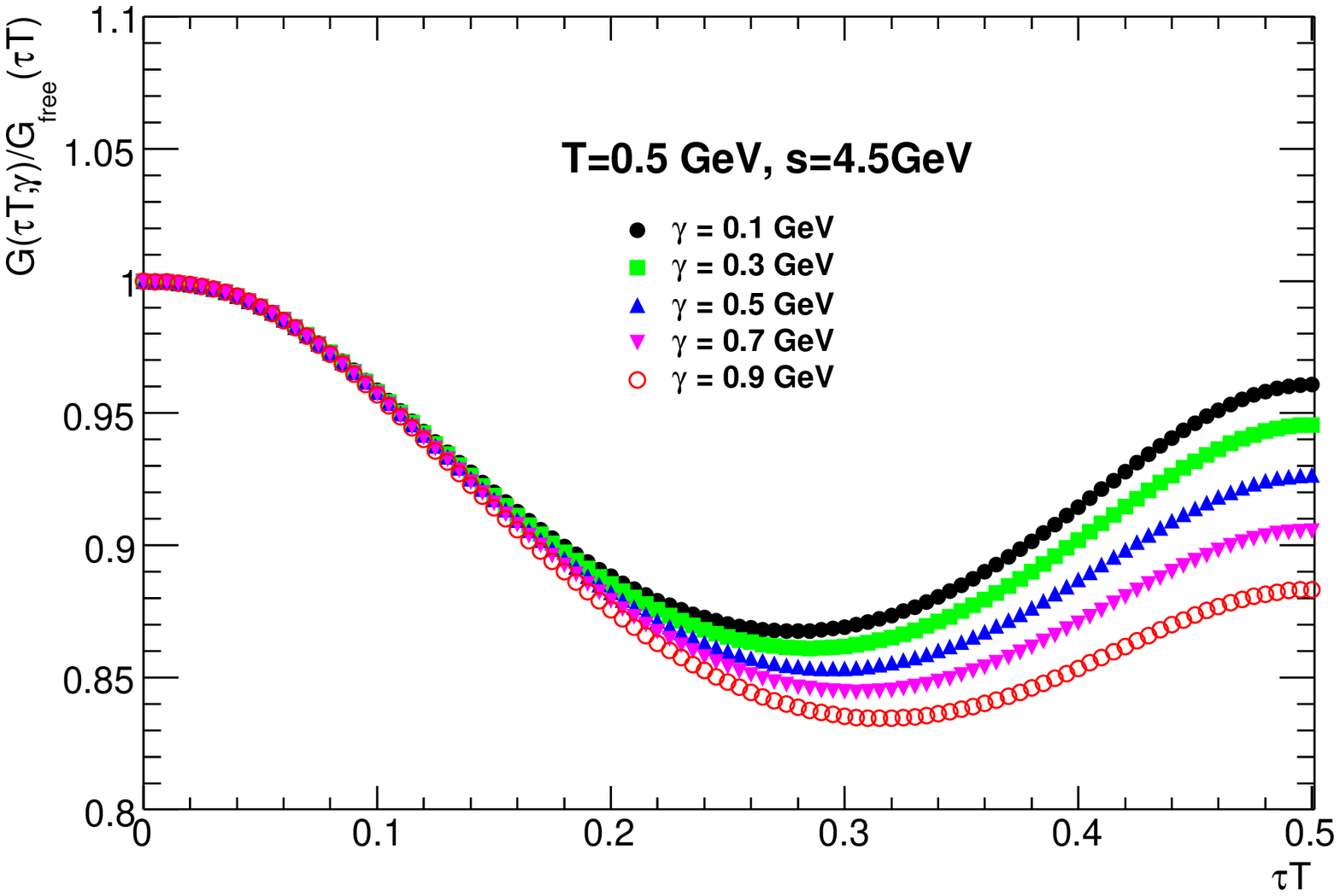,width=0.4\textwidth}
  \epsfig{file=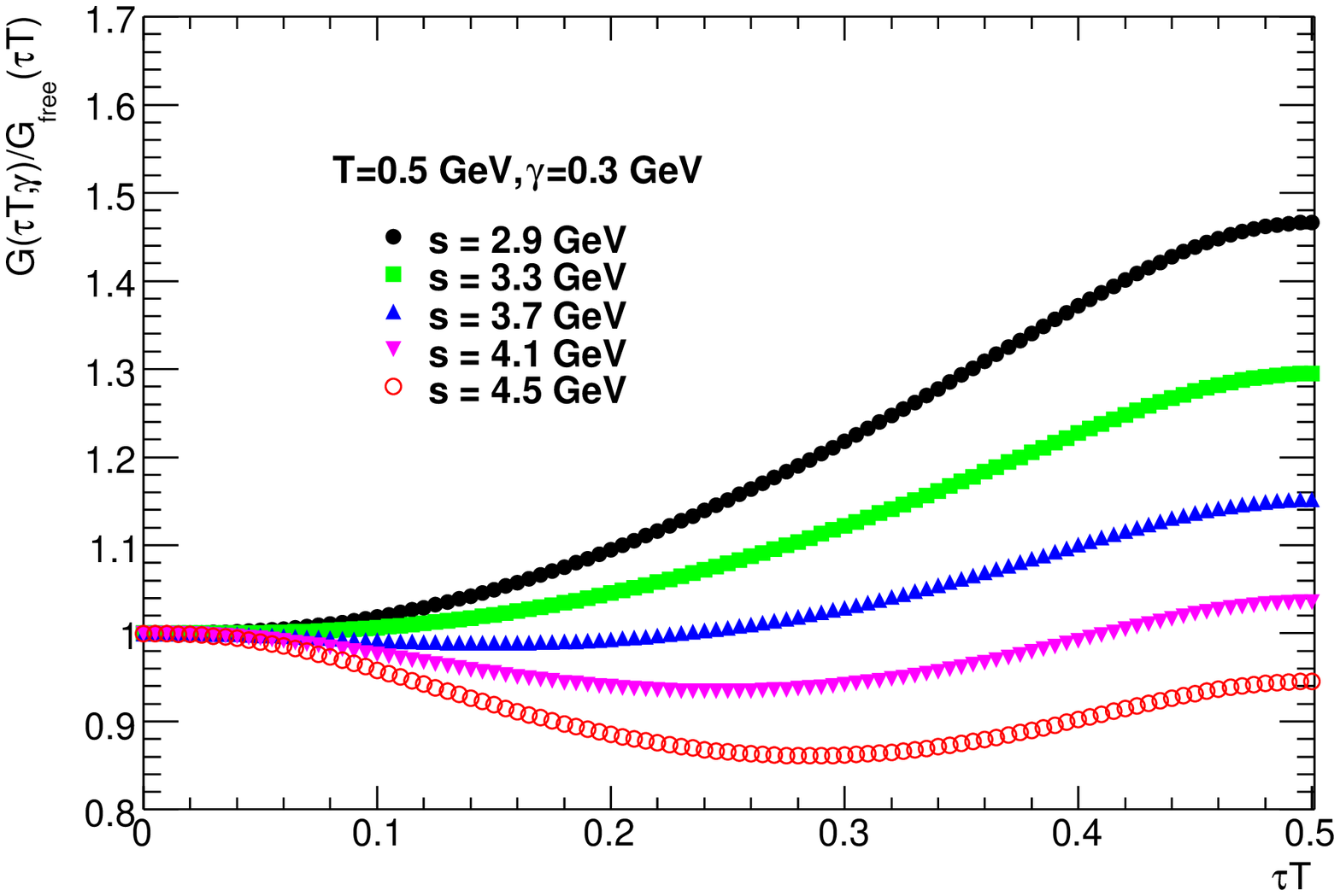,width=0.4\textwidth}
\caption{
Ratio of correlator $G(\tau,T)$ to $G_{free}(\tau,T)$ versus $\tau T$: with different width of the resonance (left) and different threshold of the continuum (right).
}
\end{figure}

\section{Reliability of MEM }

After checking the charmonium on the correlator level, we're going to the spectral function level. The normal technique to extract the spectral functions from the correlator is the MEM, by maximizing a function $Q(\sigma;\alpha)=\alpha S[\sigma] - L[\sigma]$. $L[\sigma]$ is the usual likelihood function and minimized in the standard $\chi^2$ fit. The Shannon-Jayes entropy $S[\sigma]$ is defined as
\be
S[\sigma] = \int_0^{\infty}{\mathrm{d}\omega[\sigma(\omega) - m(\omega) - \sigma(\omega)log(\frac{\sigma(\omega)}{m(\omega)})]},
\ee
where $m(\omega)$ is the default model (DM hereafter) and should be given as a plausible form of $\sigma(\omega)$. $\alpha$ is a real and positive parameter which controls the relative weight of the entropy S and the likelihood function L. The final output $\sigma_{out}$ is determined from a weighted average over $\alpha$, $\sigma_{out}\simeq \int{\mathrm{d}\alpha \sigma_{\alpha}(\omega)P[\alpha|Dm]}$,
where the most probable spectral function $\sigma_{\alpha}(\omega)$ for given $\alpha$ is obtained by maximizing the Q and $P[\alpha|Dm]$ is the weight factor.
 \begin{figure}[htp]
 \centering
 \epsfig{file=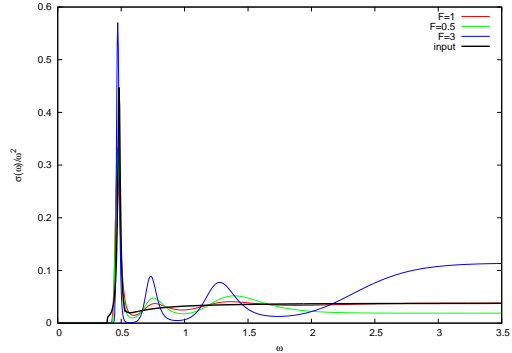,width=0.45\textwidth}
\caption{The DM dependence of spectral function obtained from MEM.}
\label{r0r1.fig}
\end{figure}
 \begin{figure}[t]
\centering
  \epsfig{file=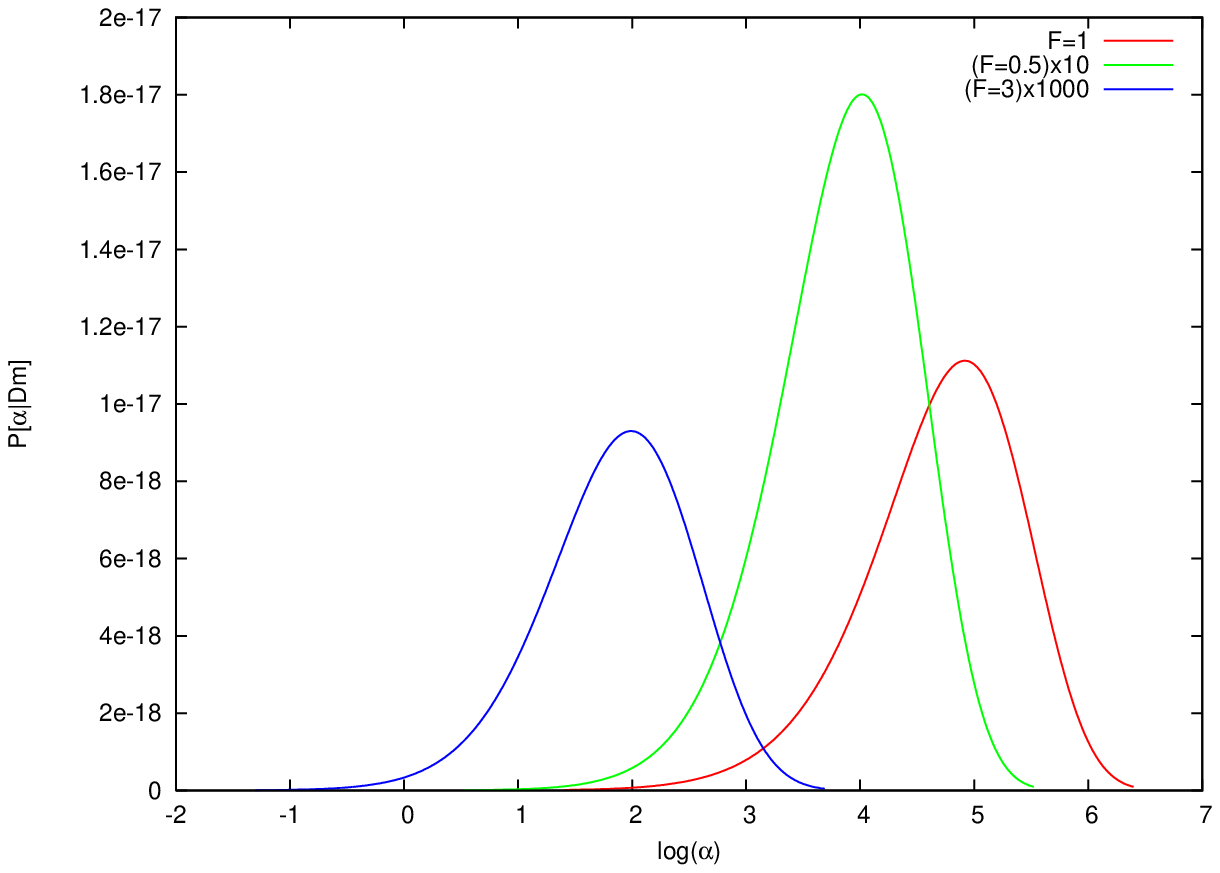,width=0.4\textwidth}
  \epsfig{file=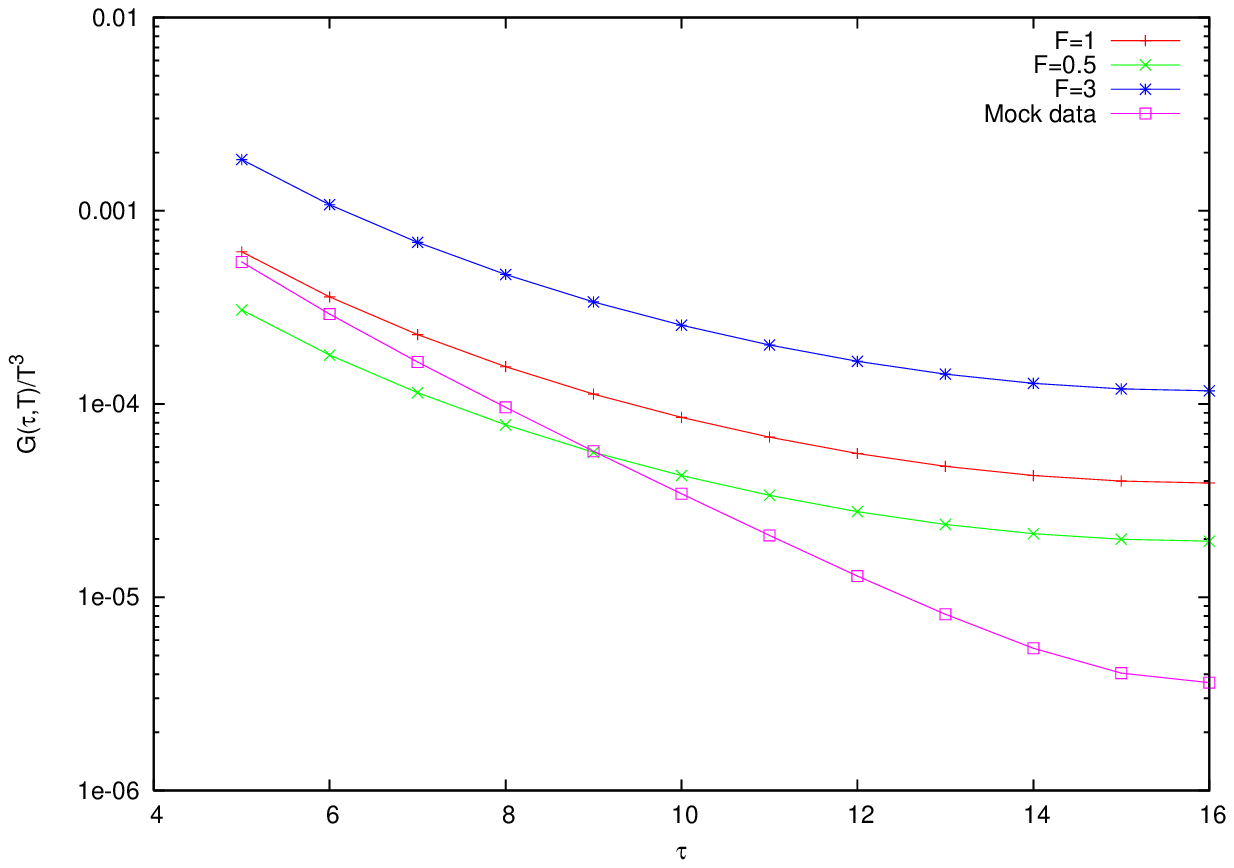,width=0.4\textwidth}
\caption{Weight factor distribution (left, for the visibility, the green line and the blue line are vertically multiplied by 10 and $10^3$, respectively), correlators calculated from the DM and the input correlators (right).}
\end{figure}
The DM is very important as it strongly affects the output of the MEM when the quality of the data is not sufficient.  As we're focusing on the modification of the ground state of the spectral function, the nature choice of the DM is the asymptotic behavior of the spectral function at large $\omega$ in the free limit. Fig. 3 shows the outputs of the MEM when using different default models ($m(\omega)=F*\frac{3}{4\pi^2}\omega^2$). The black line is the input spectral function (one resonance plus continuum) and the mock data is obtained by adding random Gaussian noises.  All the three default models reproduce the location of the resonance well, in which the one with F=1 (has the same large $\omega$ behavior as the input spectral function) gives the most reliable image. For the one with F=3, the output gets wiggled after the resonance, which is normally considered as "lattice artifacts" but could also be the "MEM artifacts"~\cite{latticeQCDstudy}.  At this point, it could be better to use the free lattice spectral function~\cite{freespf} as the DM in practice.

However, due to the lattice cutoff, such an asymptotic behavior (like DM with F=1) is not so obvious and can not be obtained directly in the lattice simulation. It could be helpful to look into the weight factor distributions and the correlators calculated from the default models, which are shown in Fig. 4. When one puts some physical prior information into the DM, it could be better that the correlators calculated from the DM is somehow comparable to the lattice correlator data (and consequently the larger peak location or amplitude of the weight factor function, see left panel of Fig. 4) rather than some orders of differences (see right panel of Fig. 4). And one also has to keep in mind that, no matter what kind of DM used, the correlators, that calculated from the output spectral functions that obtained from the MEM, can always reproduce the lattice correlator data within the errors. This essentially accents the importance of the prior knowledge of the spectral function to put into the DM and a careful analysis of the DM dependence.
 
 \section{Summary}

Within current scenario of the spectral function, the correlator is more sensitive to the change of the continuum part than the resonance part, which makes the exploration of the resonance's properties  more difficult. For the frequently used method MEM, it may also produce some artifacts. As MEM can always reproduce the lattice correlator data, it's very important to put as much physical information as possible into the DM. When one is suspicious about some parts of the spectral function from the MEM, it could be helpful to look into the correlators calculated from the default models and the weight factor distributions. In this analysis the zero mode contribution~\cite{zeromode} is not included, which requires further research.

\noindent
{\bf Acknowledgments}\\
H.-T. Ding thanks J. Engels for the assistance on the MEM.

\end{document}